\documentclass[aps,pre,twocolumn,superscriptaddress,preprintnumbers,amsmath,amssymb,floatfix]{revtex4-1}

\usepackage{grffile}
\usepackage{graphicx}
\usepackage{stackrel}

\newlength{\onefig}
\newlength{\twofig}

\usepackage[usenames,dvipsnames]{color}

\newcommand{\al}{{\alpha}}
\newcommand{\be}{{\beta}}
\newcommand{\ga}{{\gamma}}
\newcommand{\sss}{S^{(3)}}
\newcommand{\ccc}{c^{(3)}}
\newcommand{\rrr}{g^{(3)}}
\newcommand{\abc}{{\alpha\beta\gamma}}
\newcommand{\abcp}{{\alpha^\prime\beta^\prime\gamma^\prime}}

\begin{document}

\title{Static triplet correlations in glass-forming liquids: A molecular dynamics study}

\author{Daniele Coslovich}
\email{Email: daniele.coslovich@univ-montp2.fr}
\affiliation{Universit{\'e} Montpellier 2, Laboratoire Charles Coulomb UMR 5221, Montpellier, France}
\affiliation{CNRS, Laboratoire Charles Coulomb UMR 5221, Montpellier, France}

\date{\today}

\begin{abstract}
We present a numerical evaluation of the three-point static correlations functions of the Kob-Andersen Lennard-Jones binary mixture and of its purely repulsive, Weeks-Chandler-Andersen variant. In the glassy regime, the two models possess a similar pair structure, yet their dynamics differ markedly. The static triplet correlation functions $\sss$ indicate that the local ordering is more pronounced in the Lennard-Jones model, an observation consistent with its slower dynamics. A comparison of the direct triplet correlation functions $\ccc$ reveals that these structural differences are due, to a good extent, to an amplification of the small discrepancies observed at the pair level. We demonstrate the existence of a broad, positive peak at small wave-vectors and angles in $\ccc$. In this portion of $k$-space, slight, systematic differences between the models are observed, revealing ``genuine'' three-body contributions to the triplet structure. The possible role of the low-$k$ features of $\ccc$ and the implications of our results for dynamic theories of the glass transition are discussed.

\end{abstract}


\maketitle

\section{Introduction}\label{sec:intro}

The origin of the vertiginous increase of structural relaxation times of a viscous liquid cooled towards its glass transition continues to be the subject of intense research within the condensed matter community. In recent years, numerical simulations and experiments have provided evidence that the growth of relaxation times is accompanied by non-trivial dynamic correlations in the motion of the molecules at the nanometric scales, also known as dynamic heterogeneities~\cite{Berthier_Biroli_Bouchaud_Cipelletti_Saarloos_2011}. A quantitative characterization of dynamic heterogeneities requires the evaluation of four-point dynamic correlation functions, which account for spatial correlations in the fluctuations of the dynamics~\cite{Glotzer_Novikov_Schroder_2000,berthier_spontaneous_2007-1,berthier_spontaneous_2007,Berthier_Biroli_2011}. Such a description eventually provided a demonstration of the role of cooperativity in supercooled liquids, often invoked in the past to explain the increase of relaxation times~\cite{Adam_Gibbs_1965}.

Multi-point dynamic correlation functions are readily calculated from the particles' trajectories in computer simulations or in confocal microscopy measurements of colloidal suspensions. However, direct evaluation for common molecular glass-formers, such as OTP, is at present impossible. Notwithstanding, the four-point correlation functions can be approximated~\cite{berthier_direct_2005,berthier_spontaneous_2007-1,berthier_spontaneous_2007} in terms of three-point correlations by taking the temperature dependence of two-point dynamic functions, thereby allowing an indirect evaluation from  experimental data~\cite{dalle-ferrier_spatial_2007,capaccioli_dynamically_2008}. 

From the theoretical side, prediction of three-point dynamic correlation functions has been possible within the so-called inhomogeneous mode-coupling theory (IMCT)~\cite{biroli_inhomogeneous_2006}, an extension of the standard mode-coupling theory (MCT)~\cite{Gotze_2009} of the glass transition, which accounts for the dynamic response of a fluid to an external field. The predictions of IMCT suffers from the same shortcomings of MCT, i.e., the dynamic correlation volume diverges at a finite temperature; furthermore, the predicted scaling behavior in the vicinity of the avoided, putative singularity is inconsistent with the available simulation data for four-point dynamic susceptibilities~\cite{berthier_spontaneous_2007,Szamel_Flenner_2010}. 
Dynamic facilitation models~\cite{garrahan_coarse-grained_2003} explain growing dynamic correlations in terms of kinetic constraints on the local mobility. It can be argued, however, that the existence of dynamic heterogeneities in these models is ``built-in'' and therefore lacks a microscopic basis. Thus, despite significant advances in the field, a full theoretical and experimental understanding of dynamical correlations beyond the bare two-point level is still missing.

Along with dynamical correlations, some glass-forming liquids appear to develop non-trivial static correlations upon cooling below the onset temperature. Indeed, several recent studies have successfully identified static correlation length scales and studied their temperature dependence~\cite{biroli_thermodynamic_2008,Karmakar_Lerner_Procaccia_2012,charbonneau_geometrical_2012,hocky_growing_2012,berthier_static_2012}. The rationale behind these studies is that not only the dynamics, but also the local structure is heterogeneous. Such static heterogeneities may correspond to domains formed by locally preferred structures, as those envisaged in frustration-based theories of the glass transition~\cite{tarjus_viscous_2000,tanaka_two-order-parameter_2005,tanaka_two-order-parameter_2005-1}, or to the appearance of some more exotic amorphous order~\cite{montanari_rigorous_2006,biroli_thermodynamic_2008,Kurchan_Levine_2009,Berthier_Biroli_2011,Kurchan_Levine_2011}. 

Simulation studies of both simple and realistic models of metallic glass-formers have given concrete shapes to locally preferred structures~\cite{tomida_molecular-dynamics_1995,dzugutov_decoupling_2002,jain_role_2005,shintani_frustration_2006,ladadwa_low-frequency_2006,coslovich_understanding_2007-1,royall_direct_2008,Fujita_Guan_Sheng_Inoue_Sakurai_Chen_2010,hirata_direct_2011,Ding_Cheng_Sheng_Ma_2012,Sheng_Ma_Kramer_2012}, e.g., icosahedra or more general polyhedral structures. Almost invariably, such static heterogeneities are found to be connected to the dynamic properties, such as dynamic heterogeneities or fragility. Simulation~\cite{hentschel_statistical_2007,lerner_statistical_2009} and experimental studies~\cite{Mazoyer_Ebert_Maret_Keim_2011,kawasaki_structural_2011} of bidisperse or polydisperse soft particles in two dimensions reveal how the competition between hexatic order and defective local structures impact on the dynamics. Further evidence of the connection between static and dynamic length scales~\cite{sausset_growing_2010} is provided by simulations of Lennard-Jones particles confined on the hyperbolic plane. On the other hand, a related, a recent study~\cite{charbonneau_geometrical_2012} points towards a marginal role of static correlations in hard sphere models, indicating that the existence of a link between structure and dynamics may be system-specific.

Properties used to characterize static heterogeneities, such as bond-order parameters, are often tailored to detect specific types of local order. Furthermore, different geometrical constructions, e.g., Voronoi tesselations and common neighbor analysis, may lead in some cases to inconsistent results~\cite{Fang_Wang_Yao_Ding_Ho_2010}. Therefore, a more general and ``order-agnostic'' framework to evaluate static correlations is desirable. A recent line of research builds on the assumption that the relevant static correlations in glassy systems are of high order, i.e., well beyond the pair level. Indeed, Voronoi and common neighbor analysis typically involve correlations between several particles  at a time within the first neighbor shell. One general method to account for high order static correlations in a liquid is to evaluate the so-called ``point-to-set'' correlations~\cite{montanari_rigorous_2006}. This technique requires pinning a fraction of the particles and evaluating the influence of the external field of the immobile particles on the structure of the liquid~\cite{berthier_static_2012}. At least for one model glass-former, the popular Lennard-Jones Kob-Andersen (KA) mixture, point-to-set correlations have been shown to possess a non-trivial connection to the dynamics~\cite{hocky_growing_2012}. Specifically, Hocky et al. have explained the differences in dynamic behavior between two variants of the KA model (with and without attractions) in terms of different rates of growth of the point-to-set lengths. Their results are consistent with a previous analysis of the temperature variations of the local structure of these models~\cite{coslovich_locally_2011}. 

Here we take a different point of view. Instead of calculating \textit{many}-body correlations functions, without knowing exactly which order is involved, we ask ourselves whether it is possible to extract relevant information on the glassy dynamics 
already from \textit{few}-body correlation functions.
In our previous work~\cite{coslovich_locally_2011} it was found that the distribution of angles formed by nearest neighbors is more sensitive than the pair correlation functions as an indicator of the differences between attractive and repulsive variants of the KA model. In the present work, we perform a systematic evaluation of the three-point static correlations of these two models. We find that a substantial part of the difference in the triplet correlations arise from an amplification of the small differences in the pair correlations within the first coordination shell. The direct triplet correlation functions of the models differ slightly, but systematically, at larger length scales. Finally, we briefly discuss the different role of direct triplet correlations in close-packed and tetrahedral network liquids.

The manuscript is organized as follows. In Section~\ref{sec:results} we briefly describe the models and present our results. In Section~\ref{sec:conclusions} we discuss the implications for dynamic theories of the glass transition and give our conclusions.

\section{Models and results}\label{sec:results}

We report molecular dynamics (MD) simulations for two versions of the Kob-Andersen (KA) binary mixture~\cite{kob_testing_1995}: in the original model, particles interact via the Lennard-Jones (LJ) potential
\begin{equation}\label{eqn:lj} 
u_{\alpha\beta}(r) = 4\epsilon_{\alpha\beta}\left[
  {\left( \frac{\sigma_{\alpha\beta}}{r} \right)}^{12} - {\left(
    \frac{\sigma_{\alpha\beta}}{r} \right)}^6 \right]
\end{equation}
where $\alpha, \beta = 1,2$ is a species index. The values of the parameters can be found in the original paper~\cite{kob_testing_1995}. The potentials are cut and quadratically shifted~\cite{stoddard_numerical_1973} at $2.5\sigma_{\alpha\beta}$. We also study the so-called Weeks-Chandler-Andersen (WCA) variant of the model~\cite{chandler_lengthscale_2006}, in which the potentials $u_{\alpha\beta}(r)$ in Eq.~\eqref{eqn:lj} are truncated and shifted at the minimum and are therefore purely repulsive. The parameters $\sigma_{11}$, $\epsilon_{11}$, and $\sqrt{m_1\sigma_{11}^2/\epsilon_{11}}$ constitute our units of distance, energy, and time, respectively. We simulate systems of $N=1000$ particles in a cubic box with periodic boundary conditions at a constant number density $\rho=N/V=1.2$.
All simulations were carried out in the NVT ensemble using the Nos\'e-Poincar\'e thermostat~\cite{nose__2001} with a mass parameter $Q$=5.0. Some larger samples of $N=8000$ have also been simulated to check the behavior of static correlations at large length scales.

It has been pointed out that LJ and WCA models appear similar at the level of static pair correlations~\cite{berthier_role_2011} but display subtle differences in the local structure~\cite{coslovich_locally_2011}. Around its putative mode-coupling critical temperature ($T_\textrm{MCT}\approx 0.435$), the LJ model is around two orders of magnitude slower in relaxation times than the WCA model. This stark difference in dynamic behavior has been related to the different thermal rates of growth of clusters formed by locally preferred structures~\cite{coslovich_understanding_2007}, as well as to the different  growth rates of the point-to-set lengths~\cite{hocky_growing_2012}. Due to the strong similarity of their two-point structure, LJ and WCA models offer an ideal benchmark for testing the role of static multi-point correlations beyond the pair level. In the following, we will evaluate the three-point static correlation functions, with the additional aim of providing guidelines how to improve microscopic theories of the glass transition based on few-point static correlations, such as MCT or replica theory~\cite{Mézard_Parisi_1999}.

In multi-component systems, extracting a sensible physical picture from the triplet correlations is difficult due to the large number of parameters' combinations to consider~\cite{jorge_theory_2002-1,jorge_study_2002}. Binary mixtures are characterized by 8 independent triplet functions, each one being a function of, at least, three scalar variables. In order to tackle the complexity of parameters space and to simplify the presentation, we have chosen to focus only on selected triplets, $\abc$, of species. Our choice is motivated as follows. We will report results for 111 correlations, because the large particles ($\alpha=1$) constitute the majority species in the KA model; the corresponding correlations are thus expected to impact substantially the outcome of dynamic theories, e.g., through the memory kernel in MCT. In addition, we study 121 correlations, because structural motifs in binary alloys are known to be centered around the minority (solute) component~\cite{Miracle_2004,Sheng_Luo_Alamgir_Bai_Ma_2006,Yang_2012}, i.e., the small particles in the KA model; these correlations will thus be more sensitive to the details of the local structure. We have carefully inspected all species triplets to check the robustness of our conclusions. 

\begin{figure}[t]
\includegraphics[width=\onefig]{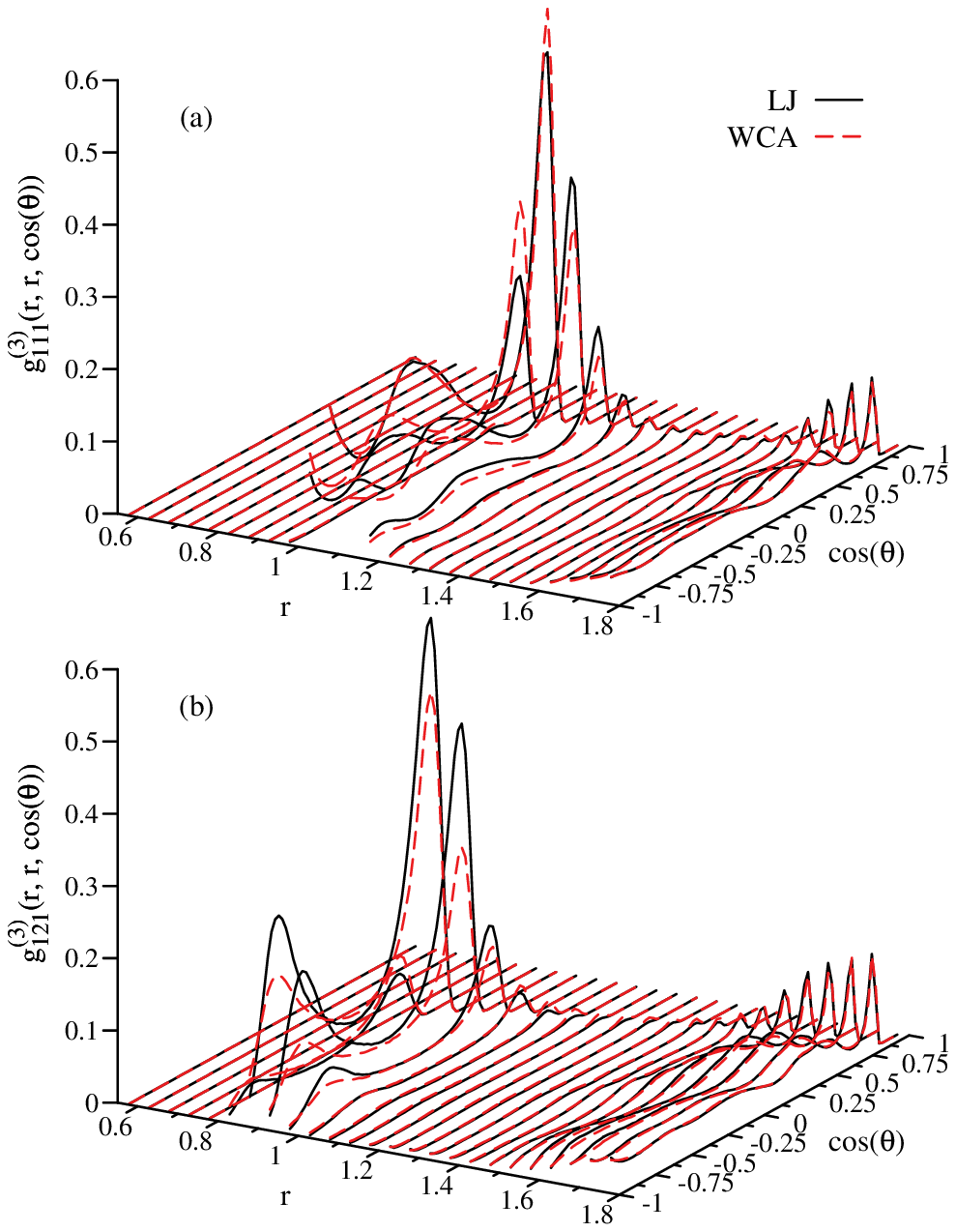}
\caption{\label{fig:g3r} Triplet correlation functions (a) $\rrr_{111}(r,r,\cos{(\theta)})$ and (b) $\rrr_{121}(r,r,\cos{(\theta)})$ for the LJ model (full lines) and the WCA model (dashed line) at $T=0.45$.}
\end{figure}

The physical content of the triplet correlations is most easily grasped in real space. Let us therefore start our presentation with a brief analysis of the triplet correlations $\rrr_\abc(\vec{r},\vec{r}^\prime)$, where $\vec{r}$ and $\vec{r}^\prime$ are the vectors connecting a central particle of species $\beta$ with two particles of species $\alpha$ and $\gamma$ at distances $\vec{r}$ and $\vec{r}^\prime$, respectively. In an isotropic, homogeneous liquid, $\rrr_{\abc}(\vec{r},\vec{r}^\prime)=\rrr_{\abc}(r, r^\prime, \cos{(\theta)})$, where $\theta$ is the angle formed by the two vectors $\vec{r}$ and $\vec{r}^\prime$. In Fig.~\ref{fig:g3r} we show $\rrr_{111}$ and $\rrr_{121}$ for the two KA models at $T=0.45$, i.e., close to the estimated $T_\textrm{MCT}$ of the LJ model. 
For simplicity, we only consider here distances along the diagonal, $r=r^\prime$. The differences between the LJ and WCA models appear more evident in the 121 than in the 111 triplet correlations. The amplitude of the peaks of $\rrr_{121}$ are in fact systematically larger in the LJ model than in the WCA. The more pronounced local ordering of the LJ model at low temperature  is consistent with its slower dynamics. By contrast, the two models display rather similar correlations between triplets of large particles. In particular, there is barely any difference in the amplitude of the main peak of $\rrr_{111}$, which arises from angles $\approx 60$ between three neighboring large particles. The peak is only shifted to slightly larger distances in the LJ model. This demonstrates that the structural differences between LJ and WCA models are mostly due to the subtle ordering of the large particles around the small ones. Thus, static properties involving correlations between unlike species are best suited to detect structural changes in the KA model, which are due to the appearance of bicapped prismatic structures~\cite{coslovich_locally_2011} centered around small particles. 
As recent work indicates~\cite{,Speck_Malins_Royall_2012}, these observations may prove useful in the analysis of the low-activity, glassy configurations generated by biasing the particle mobility~\cite{Jack_Hedges_Garrahan_Chandler_2011,speck_constrained_2012}, or of the ultra-stable glasses obtained by vapor deposition techniques~\cite{Singh_de_Pablo_2011}. Finally, we point out that the bond-angle correlation function $D_{121}(\theta)$ presented in Ref.~\cite{coslovich_locally_2011} can be obtained as the integral of $\rrr_{121}(\vec{r}, \vec{r}^\prime)$ over the first shell of neighbors.

From the point of view of microscopic dynamic theories, correlations are most conveniently represented in Fourier space.
In the following, we will therefore focus on the Fourier transform of the triplet correlation function
\begin{equation}
  \label{eq:s3k}
  \sss_\abc(\vec{k},\vec{q}) = \frac{1}{N} \langle \rho_\alpha(\vec{q}) \rho_\beta(\vec{k}) \rho^*_\gamma(\vec{k}+\vec{q}) \rangle
\end{equation}
where $\alpha,\beta$, and $\gamma$ are indices of species and $\rho_\alpha(\vec{k})$ is the Fourier transform of the microscopic density $\rho_\alpha(\vec{k})=\sum_{j=1}^{N_\alpha} \exp(i \vec{k}\cdot \vec{r}_j)$. We will often refer to the ensemble of all possible triplet correlations simply as $\sss$. Under hypothesis of homogeneity and isotropy, $\sss$ is a function of $(k,q,\cos{\theta})$, where $\theta$ is the angle formed by $\vec{k}$ and $\vec{q}$. For the sake of simplicity we will mostly concentrate ourselves on slices in Fourier space corresponding to diagonal wave-vectors, $k=q$. Our conclusions are confirmed by an analysis of the triplet correlations as a function of $k$ and $q$ at fixed angle $\theta$. 
For completeness, a brief discussion of triplet correlations for off-diagonal wave-vectors will be included at the end of the section. 

\begin{figure}[t]
\includegraphics[width=\onefig]{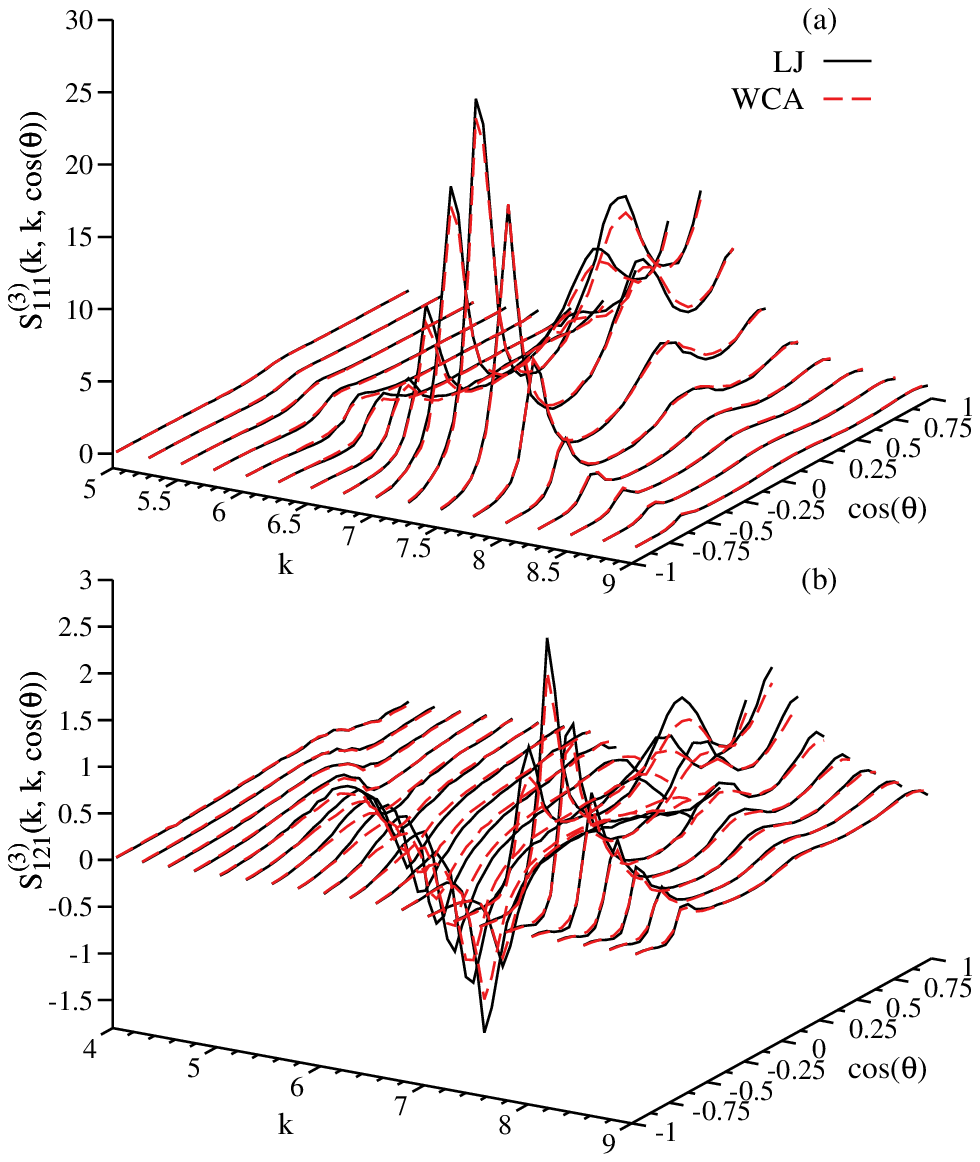}
\caption{\label{fig:s3k} Triplet correlation functions (a) $\sss_{111}(k,k,\cos{(\theta)})$ and (b) $\sss_{121}(k,k,\cos{(\theta)})$ for the LJ model (full lines) and the WCA model (dashed line) at $T=0.45$.}
\end{figure}

\begin{figure}[t]
\includegraphics[width=\onefig]{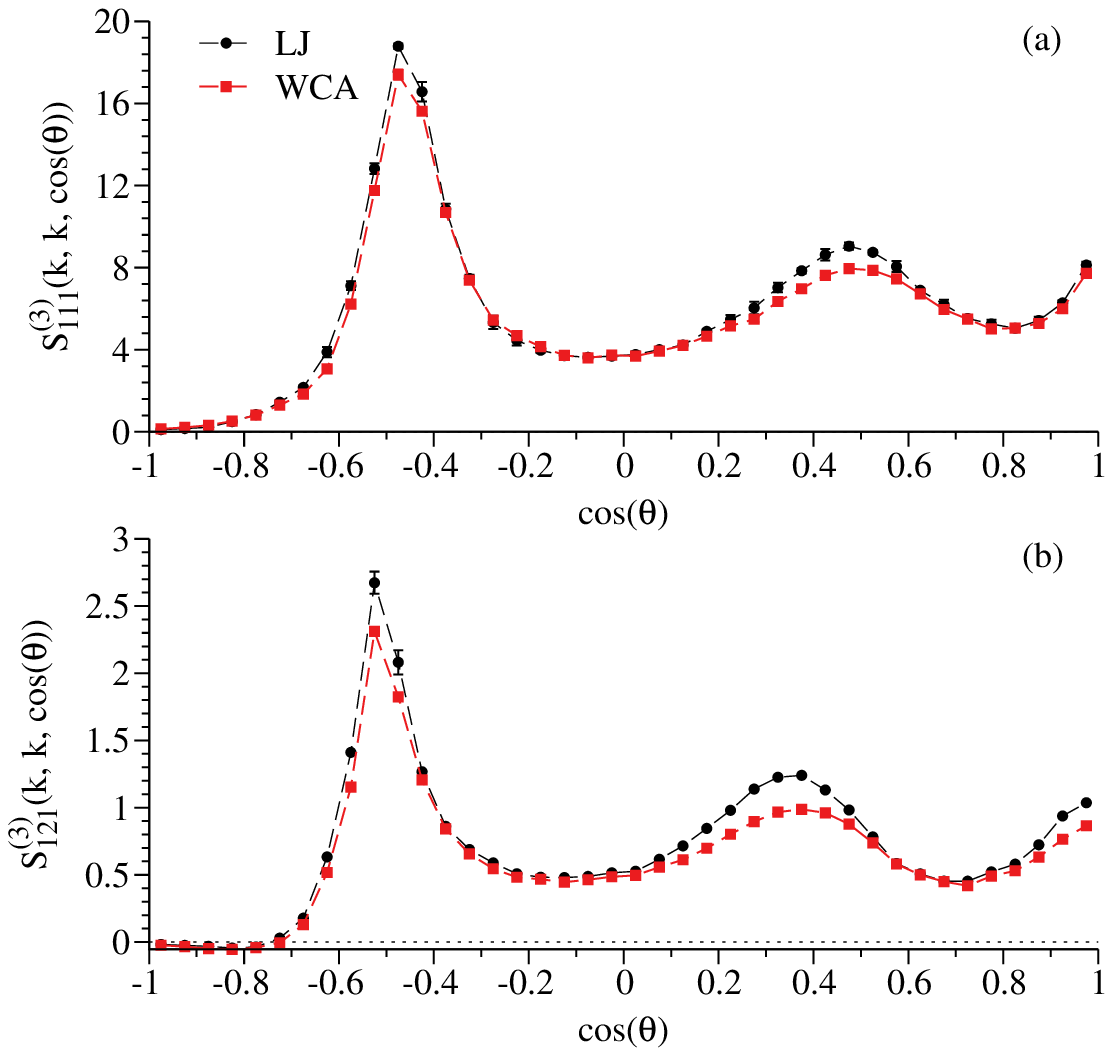}
\caption{\label{fig:s3k_kstar} Triplet correlation functions (a) $\sss_{111}(k=7,k=7,\cos{(\theta)})$ and (b) $\sss_{121}(k=7.5,k=7.5,\cos{(\theta)})$ for the LJ model (full lines) and the WCA model (dashed line) at $T=0.45$.}
\end{figure}

Figure~\ref{fig:s3k} shows the triplet correlation functions $\sss_{111}(k,k,\cos(\theta))$ and $\sss_{121}(k,k,\cos(\theta))$ evaluated at $T=0.45$. Triplet correlations between large particles display two prominent peaks along the cut $k=q=7$, for angles $\cos({\theta})\approx -0.5$ and $\cos({\theta})\approx 0.45$, respectively (see also Fig.~\ref{fig:s3k_kstar}). The dominant wave-vector ($k^*=7$) is also the location of the first peak in the structure factor $S_{11}(k)$. To pinpoint the physical origin of these ``preferred'' angles, we tentatively compare the peaks' intensities to those in real space (Fig.~\ref{fig:g3r}). For instance, the largest peak at wide angles in $\sss_{111}$ could be attributed to the preferred angle of $\theta\approx 60$ between three neighboring large particles.
In the case of $\sss_{121}$, the negative minimum around $k=6$ arises in correspondence to the first minimum of the structure factor $S_{12}(k)$. Inspection of $\sss$ as a function of $q$ and $k$ at fixed angle $\theta$, shows that $S^{(3)}_{111}$ is symmetric in $q$ and $k$, as expected, and takes its absolute maximum on the diagonal. The absolute maximum of $\sss_{121}$ is attained for slightly off-diagonal wave-vectors.

As expected, the systematic differences visible in $\sss_{121}$ confirm the more pronounced ordering of the LJ model. In particular, the amplitude (in absolute values)  of the peaks of $\sss_{121}$ is systematically larger in the LJ than in the WCA model. Differences in $\sss_{111}$ are only visible at small angles; by contrast, barely any difference is observed around the main peak. These features are further highlighted in Fig.~\ref{fig:s3k_kstar}, which shows the angular dependencies of the $\sss_{111}$ and $\sss_{121}$ at fixed wave-vector $q=k=k^*$. The different choices of $k^*$ for $\sss_{111}$ and $\sss_{121}$ reflect the locations of the meain peaks of $S_{11}$ and $S_{12}$, respectively.

\begin{figure}[t]
\includegraphics[width=\onefig]{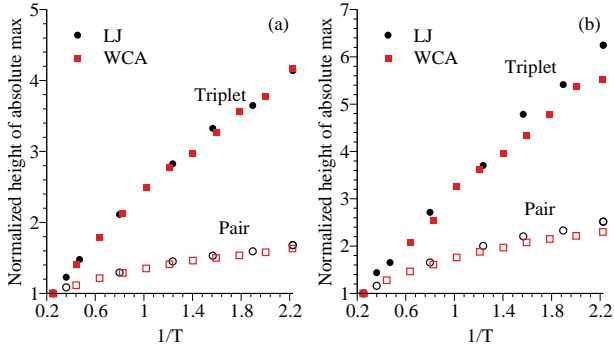}
\caption{\label{fig:maxima} Temperature dependence of the absolute maxima of the triplet correlations and the structure factors: (a) $\bar{\sss_{111}}$ (filled symbols) and $\bar{S}_{11}$ (open symbols); (b) $\bar{\sss_{121}}$ (filled symbols) and $\bar{S}_{12}$ (open symbols). The amplitudes of the maxima have been normalized by the corresponding values at $T=4$. Circles and square indicate results for LJ and WCA models, respectively. }
\end{figure}

We found that triplet correlations are fairly sensitive to temperature variations---much more than pair correlations. The locations of the peaks are only   marginally affected by temperature variations, but their amplitudes (in absolute value) show a substantial increase upon cooling. To quantify the latter effect, we identify the absolute maximum of $\sss$ over all wave-vectors $\vec{k}$ and $\vec{q}$ and track the evolution of its height $\bar{S}^{(3)}_\abc$ as a function of $T$, see Fig.~\ref{fig:maxima}. We inspected the location of the absolute maxima to make sure the peaks evolve smoothly, i.e., the geometrical meaning of the peak is consistent across temperatures.
For comparison, we also include the heights $\bar{S}_{11}$ and $\bar{S}_{12}$ of the main peaks of the structure factors $S_{11}(k)$ and $S_{12}(k)$. 
The temperature dependence of $\bar{S}^{(3)}$ is stronger than that of $\bar{S}$. The variation of the absolute maxima of $\sss$ covers a factor 6 -- 7 (depending on species triplet) going from high ($T=4$) to low ($T=0.45$) temperature. By contrast, the peaks of the structure factors increase by a more modest factor two. Further, it can be seen that, below the onset temperature ($T\approx 1.0$) of the LJ model, $\bar{S}^{(3)}_{121}$ increases at a higher thermal rate in the LJ model than in the WCA. As it may be expected, however, the difference is not as marked as the one revealed by the analysis of the locally preferred structures~\cite{coslovich_locally_2011} or by higher order correlations~\cite{hocky_growing_2012}. Of course, our simplified analysis carries no information on how the spatial extension of triplet correlations changes with temperature. Whether a growing static length can be extracted from three-point correlations is an issue that deserves further investigation.

The above results demonstrate that static correlations become more sensitive to temperature variations with increasing order. This, in turn, makes it easier to disentangle the behaviors of the LJ and WCA models. Careful inspection of the static structure factors (or the radial distribution functions) reveals, however, that small, systematic differences can be discerned already at the two-body level. It is thus natural to ask whether the differences observed in the three-point correlations arise from an ``amplification'' of those already visible at the two-body level, or reflect some ``genuine'' many-body effects. A similar issue has been recently discussed in connection to point-to-set correlations~\cite{Szamel_Flenner_2012}.

At the level of three-point static correlations, the above question can be phrased precisely in the language of liquid state theory: to what extent the convolution approximation accounts for the differences in $\sss$ between the two models? In fact, the convolution approximation factorizes the triplet correlations into products of structure factors, i.e., two-point functions; for a multi-component system it reads
\begin{eqnarray}
  \label{eq:conv}
  \sss_\abc(k,q) &\approx& S^\text{conv}_\abc(\vec{k},\vec{q}) \nonumber \\
 &= & N\rho^2 \sum_{\abcp} [ S_{\al\al^\prime}(k) S_{\be\be^\prime}(q) S_{\ga\ga^\prime}(|\vec{k}+\vec{q}|) \nonumber \\ 
 & &  \delta_{\al^\prime\be^\prime}\delta_{\al^\prime\ga^\prime}\delta_{\ga^\prime\be^\prime} ]
\end{eqnarray}
Note that, in contrast to the simple one-component case, for which
\begin{equation}
  \label{eq:conv_1}
  S^\text{conv}(k,q) = S(k) S(q) S(k+q)\,,
\end{equation}
the triplet correlation, say, $\sss_{111}$ receives now contributions from $S_{1\beta}$, with arbitrary $\beta$. The direct triplet correlation functions for a multi-component system are then defined by the linear system~\cite{barrat_equilibrium_1988}
\begin{eqnarray}
  \label{eq:c3}
  \sss_{\abc}(k,q) &=& N\rho^2 \sum_{\abcp} S_{\al\al^\prime}(k) S_{\be\be^\prime}(q) S_{\ga\ga^\prime}(|\vec{k}+\vec{q}|) \nonumber\\ 
&&[\delta_{\al^\prime\be^\prime}\delta_{\al^\prime\be^\prime}\delta_{\al^\prime\be^\prime} 
 / x_{\al^\prime} + \rho^2 \ccc_{\abc}(\vec{k},\vec{q})]
\end{eqnarray}
This somewhat complex expression hides the physical meaning of the direct triplet correlation, which is evident in the one-component case:
\begin{equation}
  \label{eq:c3_1}
  \sss(k,q) = S(k)S(q)S(k+q) [ 1 + \rho^2 \ccc(k,q) ] \,.
\end{equation}
$\ccc$ is proportional to the relative deviation of the actual triplet correlation from the convolution approximation $S^\text{conv}$. Indeed, the convolution approximations, Eq.~\ref{eq:conv} and Eq.~\ref{eq:conv_1}, are obtained by setting $\ccc \equiv 0$ in Eq.~\eqref{eq:c3} and \eqref{eq:c3_1}, respectively. It is in this sense that $\ccc$ is considered as a measure of ``genuine'' three-point contributions to the triplet correlations.

\begin{figure}[t]
\includegraphics[width=\onefig]{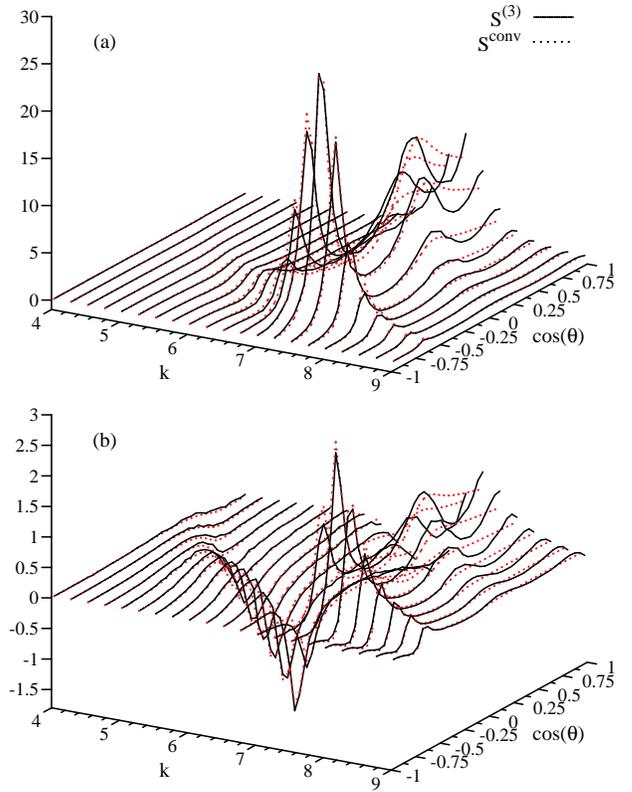}
\caption{\label{fig:s3k_conv} Triplet correlation functions $\sss(k,k,\cos{(\theta)})$ (full lines) and the convolution approximation, Eq.~\ref{eq:conv}, (dotted line) for the LJ model at $T=0.45$. (a) 111 correlations; (b) 121 correlations.}
\end{figure}

To test of the validity of the convolution approximation, we first compare $\sss$ and $S^\textrm{conv}$. From Fig.~\ref{fig:s3k_conv} we see that the convolution approximation works fairly well along the diagonal $k=q$. In particular, the major peak at wide angles in $\sss_{111}$ is accurately reproduced, and so is its temperature variation. Differences are visible at small angles, where the profiles at constant $k$  obtained by convolution appear flat and fail to reproduce the oscillations in the angular dependence. Nonetheless, an inspection of the triplet correlations for off-diagonal vectors, as well as for the WCA model, confirms that that the convolution approximation works overall reasonably well. This result may be expected in dense, hot liquids, but is non-trivial close to the estimated MCT critical temperature. These observations can be made more quantitative by analyzing the absolute errors introduced the convolution approximation (not shown here). The absolute errors are indeed largest at $k\approx k^*$ and small angles, and become negligible (typically $< 10^{-2}$) at wave-vectors smaller than $k^*$. Still the relative error at small $k$ can be very large, as discussed in the following.

\begin{figure}[t]
\includegraphics[width=\onefig]{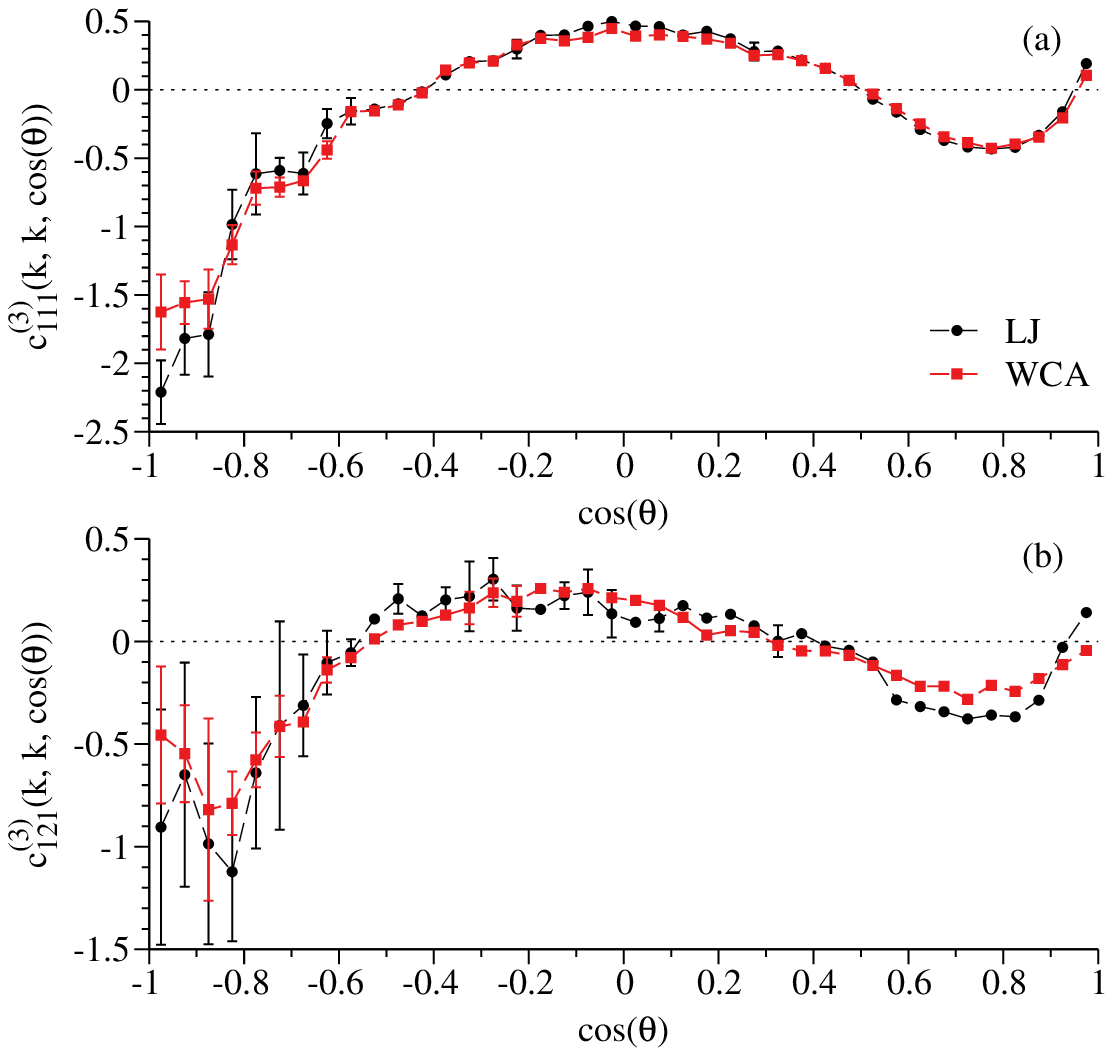}
\caption{\label{fig:c3k_star} Direct triplet correlation functions (a) $\ccc_{111}(k=7.0,k=7.0,\cos{(\theta)})$ and (b) $\ccc_{121}(k=7.5,k=7.5,\cos{(\theta)})$ for the LJ model (full lines) and the WCA model (dashed line) at $T=0.45$.}
\end{figure}

To what extent the differences in $\sss$ between the two models arise from correlations not visible at the pair level? To answer this question we evaluate the direct triplet correlations. Let us first inspect of the behavior of $\ccc_{111}$ and $\ccc_{121}$ at wave-vectors $q=k\approx k^*$ (see Fig.~\ref{fig:c3k_star}). At these wave-vectors, $\ccc$ resembles the familiar shape known from earlier studies on dense liquids with hard core interactions~\cite{barrat_factorization_1987}. Strikingly, $\ccc_{111}(k^*,k^*,\cos{(\theta)})$ matches exactly in the two models; minor differences in $\ccc_{121}(k^*,k^*,\cos{(\theta)})$ are visible at small angles, where the convolution approximation slightly underestimates correlations in the LJ model. A direct comparison of $S^\textrm{conv}$ in the two models confirm these observations. We conclude that, for diagonal wave-vectors around $k^*$, the convolution approximation introduces systematic errors in the triplet structure, but these errors are \textit{similar} in the two models. A possible physical interpretation is the following: $\ccc(k^*,k^*,\cos{(\theta)})$ is sensitive to the geometrical details of local ordering in the first neighbors shell. Since LJ and WCA models possess the same type of locally preferred structure~\cite{coslovich_locally_2011}, the direct triplet correlations around $k^*$ look very similar. The difference lies in the strength of the correlations, which is already well accounted for by the convolution approximation, and their spatial extension, which might appear at smaller wave-vectors.

\begin{figure}[t]
\includegraphics[width=\onefig]{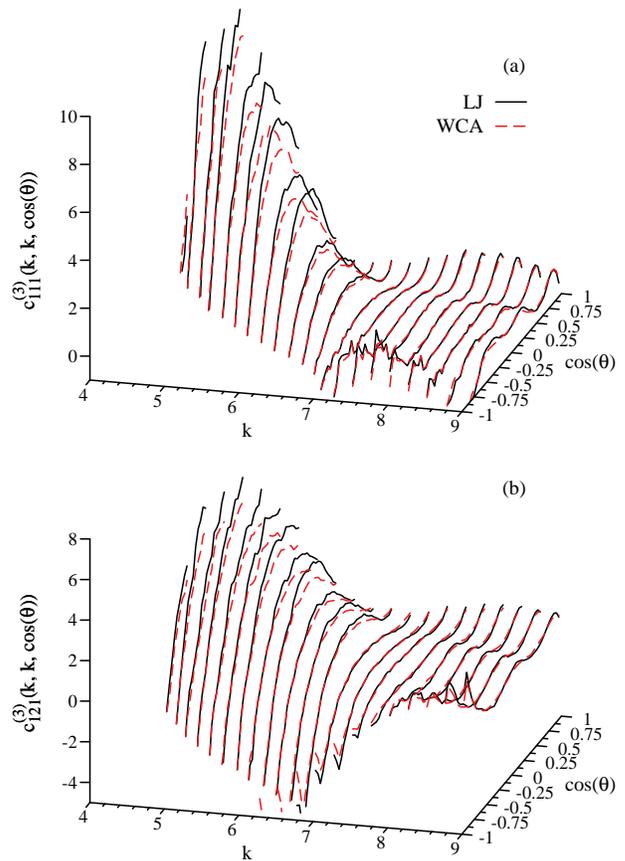}
\caption{\label{fig:c3k} Direct triplet correlation functions (a) $\ccc_{111}(k,k,\cos{(\theta)})$ and (b) $\ccc_{121}(k,k,\cos{(\theta)})$ for the LJ model (full lines) and the WCA model (dashed line) at $T=0.45$. The range on the z-axis has been limited from below for clarity.}
\end{figure}

\begin{figure*}[t]
\includegraphics[width=0.78\twofig]{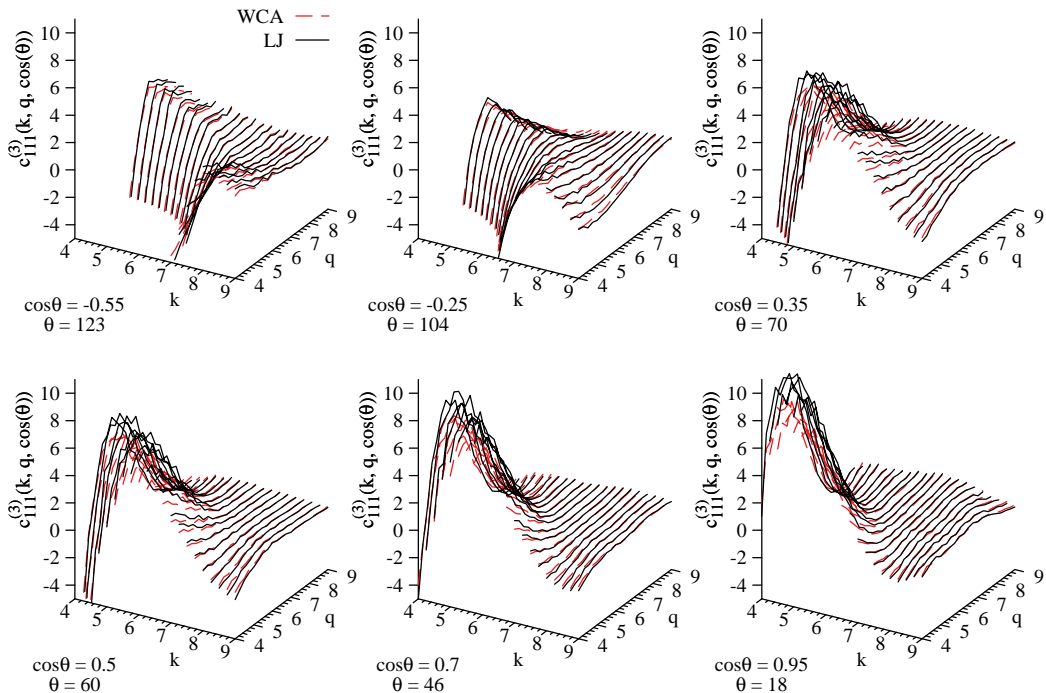}
\caption{\label{fig:c3k_off} Direct triplet correlation functions  $\ccc_{111}(k,q,\cos{(\theta)})$ and for the LJ model (full lines) and the WCA model (dashed line) at selected angles (as indicated in the labels)  at $T=0.45$.}
\end{figure*}

Inspection of $c_3$ at wave-vectors lower than $k^*$ reveals indeed small, but systematic discrepancies between the models (see Fig.~\ref{fig:c3k}). Both $\ccc_{111}(k,k,\theta)$ and $\ccc_{121}(k,k,\theta)$ develop, at small $k$ and small angles, a large positive peak, which is slightly higher in the LJ than in the WCA model. We found that the amplitude of this peak increases by decreasing temperature. Further, an investigation of the direct correlations for off-diagonal wave-vectors confirms that discrepancies in $\ccc$, whenever present below $k^*$, imply a more pronounced local ordering in LJ model than in the WCA one; see Fig.~\ref{fig:c3k_off} for representative results at selected small angles. To the best of our knowledge, the existence of such a broad maximum in the direct triplet correlation of a fluid has never been explicitly demonstrated in the literature. A positive peak in $\ccc$ at small $k$ and small angles was reported by Jorge et al.~\cite{jorge_study_2002} for an hard-sphere mixture. The amplitude of this peak was, however, very small, probably due to the relatively low packing fraction of the model employed in that study. We performed additional simulations for the hard sphere model of Ref.~\cite{jorge_study_2002} and confirmed the presence of this small maximum. At even smaller $k$, the direct correlation functions eventually become very large and negative for all species triplets, in agreement with earlier studies on hard spheres~\cite{barrat_factorization_1987,jorge_study_2002}. In this portion of $k$-space, a clear-cut comparison between the models is made difficult by the large statistical uncertainties. 

The above results indicate that at intermediate length scales there are small contributions to the triplet correlations, ignored by the convolution approximation, which reinforce the structural difference between the models. Whether the low-$k$ structure of $\ccc$ has any physical relevance is unclear at present. It can be seen from Fig.~\ref{fig:s3k} that $\sss$ is negligible in the corresponding portion of $k$ and $\theta$, and so is the absolute deviation from $S^\textrm{conv}$.  
In its standard formulation, MCT gives a strong weight to wave-vectors close to the main peaks of the static structure factors, due to the quadratic nature of the memory kernel. Thus, one may expect  contributions at small wave-vectors to be suppressed, effectively washing out any difference between LJ and WCA models beyond the pair level. 
In close-packed liquids, inclusion of $\ccc$ affects marginally the solution of the MCT equations, while a stronger effect is found in  network liquids~\cite{sciortino_debye-waller_2001}. This has been tentatively attributed in Ref.~\cite{sciortino_debye-waller_2001} to  positive contributions to the memory kernel arising from off-diagonal terms in $\ccc$. To address this issue, we also studied the triplet static correlations of a simple model of tetrahedral network liquid~\cite{coslovich_dynamics_2009}. Preliminary results indicate the existence of positive peaks at small $k$, similar to those observed for LJ and WCA models, but not for all species triplets. In the case of the network model, the small-$k$ peaks of $\ccc$ appear in the range of wave-vectors corresponding to the first sharp diffraction peak of the structure factors. This portion of $k$-space is thus still significantly weighted by the MCT memory kernel, which might explain the impact of $\ccc$ on the solution of the MCT equations for network liquid. Work to test this hypothesis is in progress.

\section{Conclusions}\label{sec:conclusions}

The comparison between LJ and WCA models has been taken by Berthier and Tarjus~\cite{berthier_nonperturbative_2009,berthier_role_2011,berthier_critical_2010,Berthier_Tarjus_2011} as a paradigmatic exemple of the limitations of microscopic theories of the glass transition based on the sole pair structure. High-order static correlations have therefore been invoked as a means to disentangle the dynamic behaviors of LJ and WCA models~\cite{coslovich_locally_2011,hocky_growing_2012}. In this work, we have analyzed the three-point static  correlations of these models. We found that the triplet correlations $\sss$, even when approximated by convolution,  are much more sensitive to temperature variations than two-point correlations. Further, we showed that differences in the triplet structure of the LJ and WCA models arise, to a good extent, from an amplification of the small discrepancies visible at the pair level. The direct triplet correlation functions $\ccc$ are in fact surprisingly similar for wave-vectors close to main peak of the structure factor. Slight, systematic differences in $\ccc$ are observed at smaller $k$, where a broad, positive peak develops for certain species triplets. Whether the low-$k$ structure of $\ccc$ may expose a static length scale relevant  for the glass transition is a question deserving further study.

The results of this study help clarifying the role of the approximations involved in dynamic theories of the glass transition based on few-point static correlations, such as MCT. One should bare in mind, for instance, that both three- and four-point static correlations are indeed taken into account by MCT, although in a factorized form. Using the pair structure factors as input, a solution of the MCT equation~\cite{berthier_critical_2010} only partly accounts for the dynamic differences between the LJ and WCA models. Given the strong similarity of their $\ccc$ and the negligible effect of $\ccc$ on the MCT solutions for close-packed liquids~\cite{sciortino_debye-waller_2001}, the deficiencies of the theory will not be cured by inclusion of the direct triplet correlation functions. Alternative approximation schemes for the memory kernel~\cite{Szamel_2004}, or different microscopic approaches, such as replica theory~\cite{Jacquin_Zamponi_2012}, may be able to amplify the small-$k$ features of $\ccc$ and thus prove more effective than the standard MCT in disentangling the dynamics of the two models. On a more general note, our work indicates that systematic inclusion of factorized static and dynamic correlations beyond the pair level, such as in generalized versions of MCT~\cite{mayer_cooperativity_2006,Szamel_2004}, is a promising line of research to improve constructively existing dynamic theories of the glass transition. 

\begin{acknowledgements} 

We acknowledge the HPC@LR Center of Competence in High-Performance Computing of Languedoc-Roussillon (France) for allocation of CPU time.
We thank L. Berthier, H. Jacquin, A. Ikeda, and W. Kob for valuable discussions.

\end{acknowledgements}

%


\end{document}